\documentclass[prd,nofootinbib,a4paper,showpacs,showkeys,preprintnumbers]{revtex4}

\usepackage[T1]{fontenc}
\usepackage[utf8]{inputenc}
\usepackage{geometry}
\usepackage{amsmath}
\usepackage{makeidx}
\usepackage{braket}
\usepackage{vector}
\usepackage{slashed}
\usepackage{mathrsfs}
\usepackage{color}

\usepackage{pstricks}

\makeindex
\usepackage{subfigure}
\usepackage{amssymb, amsmath}
\usepackage{ae}
\usepackage{times}
\usepackage[colorlinks=true,linkcolor=blue,urlcolor=blue,citecolor=blue,bookmarks=true,bookmarksopen=true]{hyperref}

\usepackage{graphicx,epsfig}
\makeatother

\newcommand{\dend}{{\, .}}
\newcommand{\kend}{{\, ,}}

\newcommand{\ampabs}[2]{\abs{\mathcal{M}_0}_{{#1}\rightarrow{#2}}}

\newcommand{\lorentzd}[2]{\dif\Pi^{#1}_{#2}}

\newcommand{\dif}{\,d}

\newcommand{\bbar}{\bar{b}}

\newcommand{\higgs}{\phi}

\newcommand{\lepton}{{\ell}}

\newcommand{\Neutrino}{{N}}

\newcommand{\tr}{\mbox{Tr}}

\providecommand{\hubblerate}{H}

\providecommand{\lorentzd}[2]{\dif\Pi^{#1}_{#2}}

\definecolor{blue}{rgb}{0,0,1}
\definecolor{red}{rgb}{1,0,0}
\definecolor{red}{rgb}{1,0,0}
\definecolor{orange}{rgb}{1,0.4,0}
\definecolor{green}{rgb}{0,1,0}
\definecolor{grey}{rgb}{0.5,0.5,0.5}

\providecommand{\eqn}{eqn.\,}
\providecommand{\eqns}{eqns.\,}

\providecommand{\dend}{{\, .}}
\providecommand{\kend}{{\, ,}}

\providecommand{\ifeqthenelse}[4]{\edef\tempa{#1}\def\tempb{#2}\ifx\tempa\tempb {#3} \else {#4}\fi}

\providecommand{\bigformula}[2]{\begin{equation}#1\ifeqthenelse{#2}{}{\nonumber}{\label{eqn:#2}} \end{equation}}

\providecommand{\bigformulalign}[2]{\begin{align}#1\ifeqthenelse{#2}{}{\nonumber}{\label{eqn:#2}}\end{align}}

\providecommand{\eqnref}[1]{(\ref{eqn:#1})}

\def\C{C}
\def\CP{CP}

\providecommand{\fig}{fig.\,}

\providecommand{\figref}[1]{\ref{fig:#1}}

\definecolor{darkgreen}{rgb}{0.0,0.3,0.0}

\providecommand{\abs}[1]{\left|#1\right|}
\providecommand{\heaviside}[1]{\Theta\ifeqthenelse{#1}{}{}{\left(#1\right)}}

\providecommand{\trash}[1]{}

\providecommand{\dif}{\,d}

\providecommand{\bbar}{\bar{b}}

\newcommand{\f}[2]{f_{#2}^{#1}}
\newcommand{\feq}[2]{f_{#2}^{#1,eq}}

\def\BES{Boltzmann equations}

\newcommand{\momk}{k}
\newcommand{\momp}{p}
\newcommand{\momq}{q}

\newcommand{\m}[1]{m_{#1}}
\newcommand{\e}[2]{E^{#1}_{#2}}

\newcommand{\lioulong}[3]{\ifeqthenelse{}{#2}{\ifeqthenelse{}{#3}{L[#1]}{L[#1](#2,#3)}}{\ifeqthenelse{}{#3}{L[#1]}{L[#1](#2,#3)}}}

\def\eprinttmp@#1arXiv:#2 [#3]#4@{
	\ifthenelse{\equal{#3}{x}}{\href{http://arxiv.org/abs/#1}{\texttt{#1}}}{\href{http://arxiv.org/abs/#1}{\texttt{#1}.}}
}
\newcommand{\eprint}[2]{\eprinttmp@#1arXiv: [y]@}

\usepackage{color}
\definecolor{blue}{rgb}{0,0,0.5}
\definecolor{lightblue}{rgb}{0,0,1}
\definecolor{red}{rgb}{1.0,0,0}
\definecolor{lightred}{rgb}{1,0.5,0}
\definecolor{green}{rgb}{0,0.5,0}
\definecolor{darkgreen}{rgb}{0.0,0.3,0.0}
\definecolor{grey}{rgb}{0.5,0.5,0.5}

\renewcommand{\Im}{{\rm Im}}

\providecommand{\momk}{{p}_1}
\providecommand{\momp}{{p}_2}
\providecommand{\momq}{{p}_3}

\providecommand{\momkabs}{|\bvec{k}|}
\providecommand{\mompabs}{|\bvec{p}|}
\providecommand{\momqabs}{|\bvec{q}|}

\providecommand{\lorig}{L}

\providecommand{\qstat}[1]{[1+#1]}
\providecommand{\qstatferm}[1]{[1-#1]}

\providecommand{\hubblerate}{{H}}

\newcommand{\carrayorigkb}[3]{{C}_{#1}\ifeqthenelse{}{#3}{}{[{#3}]}{\ifeqthenelse{}{#2}{}{(#2)}}}

\newcommand{\mpsii}{M_i}
\newcommand{\mpsij}{M_j}

\begin{document}


\preprint{TUM-HEP-749/10}

\title{Medium corrections to the \textit{CP}-violating parameter in leptogenesis
 }

\author{M. Garny$^{b}$}
\email[\,]{mathias.garny@ph.tum.de}

\author{A. Hohenegger$^{a}$}
\email[\,]{andreas.hohenegger@mpi-hd.mpg.de}

\author{A. Kartavtsev$^{a}$}
\email[\,]{alexander.kartavtsev@mpi-hd.mpg.de}

\affiliation{%
\vskip 2mm
$^a$Max-Planck-Institut f\"ur Kernphysik, Saupfercheckweg 1, 69117 Heidelberg,
Germany\\
$^b$Technische Universit\"at M\"unchen, James-Franck-Stra\ss e, 85748 Garching,
Germany}

\begin{abstract}
In two recent papers, arXiv:0909.1559 and arXiv:0911.4122, it has been
demonstrated that one can obtain quantum corrected Boltzmann kinetic equations 
for leptogenesis using a top-down approach based on the
Schwinger--Keldysh/Kadanoff--Baym formalism. These ``Boltzmann-like'' equations
are similar to the ones obtained in the conventional bottom-up approach but
differ in important details. In particular  there is a discrepancy between the
\CP-violating parameter obtained in the first-principle derivation and  in the
framework of thermal field theory.   Here we demonstrate that the two approaches
can be reconciled if \textit{causal} $n$-point functions are used in the thermal
field theory approach. The new result for the medium correction to the
\CP-violating parameter  is  \textit{qualitatively} different from  the conventional one. The
analogy to a toy model considered earlier enables us to write down consistent
quantum corrected Boltzmann equations for thermal leptogenesis in the
SM+$3\nu_R$ which include quantum statistical terms and medium corrected
expressions for the \CP-violating parameter.
\end{abstract}

\maketitle

\newcommand{\cpviol}[3]{\epsilon_{#1}^{#2,#3}}

\section{Introduction}

To calculate the baryon asymmetry generated during the epoch of 
leptogenesis \cite{Fukugita:1986hr} in the 
standard model extended by three right-handed neutrinos (SM+$3\nu_R$)  and its
extensions one usually uses standard Boltzmann kinetic 
equations. The collision terms (and in particular the \CP-violating parameters) 
in this equations are computed in vacuum in the in-out 
formalism \cite{Giudice:2004npb685,Davidson:pr2008} and do not take into
account 
effects induced by the hot medium of the early universe. Such effects 
can be consistently taken into account in a top-down approach based on the
Schwinger--Keldysh/Kadanoff--Baym formalism. In \cite{Garny:2009rv,Garny:2009qn}
we have applied it to a simple toy model of leptogenesis  and derived a new 
(quantum corrected) form of the Boltzmann equations, which includes quantum 
statistical factors and takes the medium effects into account. We have found 
that the medium corrections to the \CP-violating  parameter $\epsilon$ depend 
only linearly on the one-particle distribution functions (see also 
\cite{Kiessig:2009cm,Buchmuller:2000nd,Anisimov:2010aq}). In the analysis based on 
finite temperature field theory for the phenomenological scenario of thermal 
leptogenesis \cite{Giudice:2004npb685,Davidson:pr2008,Covi:1998prd57} and  for 
GUT baryogenesis \cite{Takahashi:1984} the medium corrections to the 
\CP-violating parameter $\epsilon$ depend quadratically  on the
distribution functions. 

This discrepancy has  been noted in the context of leptogenesis in
\cite{Garny:2009rv} for the vertex contribution to the \CP-violating parameter
and later in \cite{Garny:2009qn} for the self-energy contribution. Here, we use
a finite temperature equivalent of the Cutkosky cutting rules
\cite{Kobes1985np260p3,Kobes:1986npb272,Bedaque:1996mpl,Gelis:1997npb508} to
derive thermal corrections to the expression for the imaginary part of the
three-point vertex function and the self-energy loop  and to calculate the
corresponding medium-corrected \CP-violating parameters. We show that the
discrepancy is due to an ambiguity in the real-time (RTF) formulation of thermal
quantum field theory and disappears if one considers \textit{retarded} or
\textit{advanced} $n$-point functions. In the framework of the toy model this
has been demonstrated recently in  \cite{Hohenegger:2009phd}. Together with the
new form of the Boltzmann equation derived in \cite{Garny:2009rv,Garny:2009qn}
this puts us in the position to write down quantum corrected {\BES} for the
phenomenological scenario of thermal leptogenesis which consistently include the
medium corrected \CP-violating parameter and quantum statistical terms. 

In section \ref{sec:TQFT}, we introduce our notations for the \CP-violating
parameters
and the thermal field theory formalism. Then, in section \ref{PhysGhost} we
review the conventional
calculation of the thermal corrections, and in section \ref{Causal} we
demonstrate how to reconcile them
with the recent results from nonequilibrium field theory. Finally, in section
\ref{BoltzmannEq} we present
the quantum corrected Boltzmann equations taking medium corrections into
account.

\section{\label{sec:TQFT}\CP-violating parameter and thermal field theory}

\begin{figure}[th]
	\centering
\includegraphics{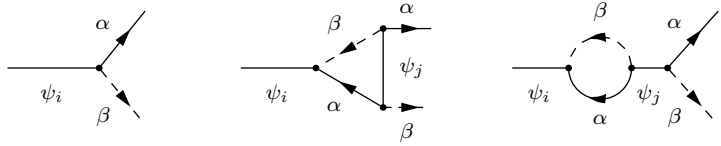}
	\caption{Tree level and one-loop contributions to the heavy Majorana neutrino
decay $\psi_i\rightarrow  \alpha\beta$. The asymmetry, at lowest order, is due
to the interference of these contributions.\label{fig:majo_decays}}
\end{figure}

In the phenomenological scenario of thermal leptogenesis as well as in the toy
model, the matter-antimatter asymmetry is generated by the decay of a heavy
species. In both cases the {\CP} violation in this decay is caused by
interference between the tree level and the one-loop diagrams, see
\fig\ref{fig:majo_decays}. 
In the phenomenological scenario $\psi_i =\Neutrino_i$ are heavy Majorana
neutrinos which decay via Yukawa interactions ${\cal L}=h_{\alpha i}\ell_\alpha
N_i\, \phi+h.c.$  into leptons $\alpha =\lepton$ and Higgs $\beta =\higgs$ or
their anti-particles. In the toy model $\psi_i$ is a heavy real scalar particle
which decays via  Yukawa interactions ${\cal L}=-\frac{g_i}{2!}\psi_i bb+h.c.$
into two light scalars $\alpha =\beta =b$ or the conjugate $\bbar$. The
\CP-violating parameter $\epsilon_i$ for the decay of $\psi_i$ is defined as 
\bigformulalign{
	\epsilon_i&=\epsilon_i^V +\epsilon_i^S =\frac{\Gamma_{\psi_i\rightarrow
\alpha\beta}
-\Gamma_{\psi_i\rightarrow\bar{\alpha}\bar{\beta}}}{\Gamma_{\psi_i\rightarrow
\alpha\beta}+\Gamma_{\psi_i\rightarrow\bar{\alpha}\bar{\beta}}}\kend
}{CP-violating parameter}
where $\Gamma_{N_i\rightarrow {\lepton}{\higgs}}$ includes a sum over flavour
indices  and loop-internal Majorana neutrino generations in the case of the
phenomenological scenario: $\Gamma_{N_i\rightarrow
{\lepton}{\higgs}}=\sum_{\alpha,\,j} \Gamma_{N_i\rightarrow
{\lepton_\alpha}{\higgs}}$ (we do not consider flavor effects here). 

\begin{figure}[!ht]
	\centering
\includegraphics{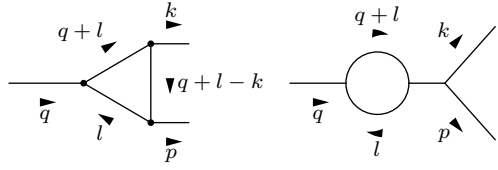}
	\caption{Momentum flow in the vertex and the self-energy
loop.\label{fig:vertex corrections momentum flow}}
\end{figure}

If the tree level and one-loop contributions are written as $\lambda_0
\mathcal{A}_0$ and $\lambda_1 \mathcal{A}_1$, respectively, where all coupling
constants are absorbed in $\lambda_{0(1)}$, the \CP-violating parameter becomes
at lowest order:
\bigformula{
	\epsilon_i=\frac{\abs{\lambda_0
\mathcal{A}_0+\lambda_1\mathcal{A}_1}^2-\abs{\lambda_0^*\mathcal{A}
_0+\lambda_1^*\mathcal{A}_1}^2}{\abs{\lambda_0\mathcal{A}_0+\lambda_1\mathcal{A}
_1}^2+\abs{\lambda_0^*\mathcal{A}_0+\lambda_1^*\mathcal{A}_1}^2}\simeq
-2\frac{\Im\big\{\lambda_0^*\lambda_1\big\}\Im\left\{\mathcal{A}_0^*\mathcal{A}
_1\right\}}{{\abs{\lambda_0}^2}\abs{\mathcal{A}_0}^2}\kend
}{cp asymmetry in terms of imaginary parts}where the sum over lepton flavour and
Majorana neutrino generation indices is again implicit.
In the case of thermal leptogenesis this leads to 
\bigformula{
	\epsilon_i=-2\sum_{j\neq
i}\frac{\Im\big\{(h^{\dagger}h)_{ij}^2\big\}}{(h^{\dagger}h)_{ii}}\frac{
\Im\left\{\mathcal{A}_0^*\mathcal{A}_1\right\}}{2\, q\cdot k}\kend\quad
i=1,2,3\kend
}{cp asymmetry in terms of imaginary parts phenomenological}
where $q$ and $k$ denote the four-momenta of the Majorana neutrino and the
lepton respectively, see \fig\figref{vertex corrections momentum flow}, and for
the toy model to
\bigformula{
	\epsilon_i=-2\abs{g_j}^2 \Im\bigg(\frac{g_i g_j^*}{g_i^*
g_j}\bigg)\Im\left\{\mathcal{A}_0^* \mathcal{A}_1\right\}\kend\quad i\neq j
\kend \,\quad i=1,2\dend
}{cp asymmetry in terms of imaginary parts toy}
This means that one needs to compute the imaginary (absorptive) part of the
vertex and the self-energy loop contributions
$\Im\left\{\mathcal{A}_0^*\mathcal{A}_1\right\}$. In vacuum this can be done
conveniently with help of the Cutkosky cutting rules
\cite{cutkosky:429,Eden:2002,Bellac:1992}. In thermal quantum field theory these
can be generalized in order to take into account interactions of internal lines
in the loops with the background medium
\cite{Kobes1985np260p3,Bedaque:1996mpl,Gelis:1997npb508}. In the real-time
formalism of thermal quantum field theory two types of fields, termed {\em
type-1} and {\em type-2} fields, are introduced in order to avoid pathological
singularities \cite{Rivers:1988}. 
Vertices can be of either type, differing only by a relative
minus sign. We denote them by $g^1 = -ig$ and $g^2 = +ig$ for a generic
coupling\footnote{At the end of the calculation of
$\Im\left\{\mathcal{A}_0^*\mathcal{A}_1\right\}$ we set $g=1$,  since the
physical coupling constants have been factored out into $\lambda_0$ and
$\lambda_1$.} $g$.
The propagators connecting the different types of vertices can be considered as
components of a $2\times 2$ propagator matrix\footnote{In
\cite{Giudice:2004npb685} and elsewhere resummed propagators  have been used in
this place to prevent the appearance of singularities. Since we
are mainly interested in the structure of the thermal corrections we stick to
the free thermal propagators here.}
\bigformula{
	G_{a}(p)=
	\left(\begin{array}{cc}
		G_a^{11}(p) & G_a^{12}(p)\\
		G_a^{21}(p) & G_a^{22}(p)
	\end{array}\right)
	= 
	\left(\begin{array}{cc}
		\Delta_a (p) & e^{\beta p_0 /2}\Delta_a^- (p) \\
		 e^{-\beta p_0 /2}\Delta_a^+ (p) & \Delta_a^* (p)
	\end{array}\right)
\dend}{}
For a scalar particle $b$ the components are
\bigformulalign{
	\Delta_b (p) = D_b (p)\kend\quad
	\Delta_b^{\pm} (p) = D_b^{\pm}(p)\dend
}{thermal Feynman rules scalars}
For a fermion $f$ the components are
\bigformulalign{
	\Delta_f (p) = (\gamma\cdot p +\m{f}) D_f (p)\kend\quad
	\Delta_f^{\pm} (p) =(\gamma\cdot p +\m{f}) D_f^{\pm}(p)\dend
}{thermal Feynman rules fermions}
For brevity we have defined
\bigformulalign{
	D_a (p) & = \frac{i}{p^2 -\m{a}^2 +i\epsilon} - 2\pi \xi_a
\feq{a}{}(p)\delta{(p^2 -\m{a}^2)}\kend\nonumber\\
	D_a^{\pm}(p) & = 2\pi \left[\heaviside{\pm p_0} - \xi_a
\feq{a}{}(p)\right]\delta (p^2 -\m{a}^2)\kend
}{thermal Feynman rules D}
where $\xi_a = +1$ for fermions and $\xi_a = -1$ for bosons.
Here, we denote by $\feq{b}{}(p)$ and $\feq{f}{}(p)$ the equilibrium
distribution function for bosons and fermions, respectively, given by
\begin{equation}
	\feq{a}{}(p) = \left[ \exp\left(\beta |p_\mu U^\mu| \right) + \xi_a
\right]^{-1} \;.
\end{equation}
 They are functions of the Lorentz invariant product $p_\mu U^\mu$ of the
particles' four-momentum and the four-velocity $U$ of the plasma in a general
frame. In the rest-frame of the plasma, $U=(1,0,0,0)$, we obtain the standard
form which depends on $p_0$.
In the following we assume that it is sufficient to replace the different
propagators in our toy model and the phenomenological theory by their thermal
field theory equivalents given in \eqn\eqnref{thermal Feynman rules scalars}.
This approach has been followed in previous works for the baryogenesis and
leptogenesis scenarios
\cite{Giudice:2004npb685,Davidson:pr2008,Covi:1998prd57,Takahashi:1984}. We
ignore further thermal effects, such as thermal corrections to the masses and
wave function renormalization here.

Denoting vertices attached to external lines by $x_i$ and those attached to
internal lines only by $z_j$ we can formally denote an amputated $n$-point graph
by $F(x_1,\ldots ,x_n;z_j)$. Here we assume that $F$ is given in momentum space,
writing the position space coordinates in order to identify the individual
vertices. The contribution of this graph to the amplitude is $-iF(x_1,\ldots
,x_n;z_j)$. 

\begin{figure}[!ht]
	\centering
\includegraphics{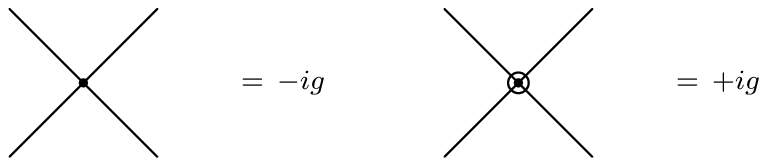}\\
	\vspace{10mm}
\includegraphics{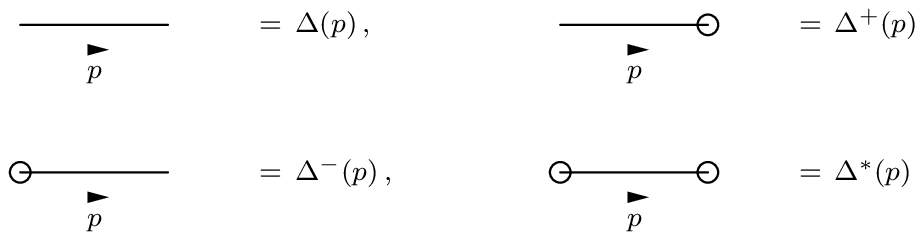}
		\caption{Circling rules for a generic theory used for the computation of
$F_{>}$ in momentum space. The rules for the computation of $F_{<}$ differ by
interchange of $\Delta^+ (p)$ and $\Delta^- (p)$. The $\Delta^{\pm}$ propagators
connecting circled and uncircled vertices may be interpreted as cut propagators.
In vacuum they correspond to the cut propagators in the Cutkosky
rules.\label{fig:circling rules}}
\end{figure}

Physical amplitudes involve a sum over possible combinations of types of
internal vertices:
\bigformula{
	\mathcal{F}(x_1,\ldots ,x_n;z_j) = \sum_{\text{type}\,z_j} F(x_1,\ldots
,x_n;z_j)\dend
}{}
For external vertices of fixed type it has been shown
\cite{Kobes1985np260p3,Kobes:1986npb272} that this sum is equivalent to a sum
over all possible ``circlings'' of the internal vertices:\footnote{The historic
origin of this formula was that the external fields where considered to be all
of type 1 (physical).}
\bigformulalign{
	\mathcal{F}(x_1,\ldots ,x_n;z_j) = & \sum_{\text{circling}\, z_j}
F_{\gtrless}(x_1,\ldots ,x_n;z_j)\dend
}{circling with fixed type of external vertices}
$F_{>}$ and $F_{<}$ with ``circled'' vertices represent graphs computed using
the set of rules, given in \fig\figref{circling rules}. These differ for the
computation of $F_{>}$ and $F_{<}$ by interchange of the $\Delta^+$ and
$\Delta^-$ propagators. In $F_{\gtrless}(x_1,\ldots ,x_n;z_j)$ we explicitly
denote circling of a vertex $\alpha$ as $F_{\gtrless}(x_1,\ldots
,\underline{x}_{\alpha},\ldots ,x_n;z_j)$. Note that the two ways of defining
$\mathcal{F}$ in terms of $F_{>}$ and $F_{<}$ in \eqn\eqnref{circling with fixed
type of external vertices} are in agreement only if the Kubo--Martin--Schwinger
(KMS) boundary condition, 
\begin{equation}
	\Delta_a^- (p) = -\xi_a e^{ - \beta p\cdot U}\Delta_a^+ (p) \;,
\end{equation}
is satisfied. This is the case in thermal equilibrium.

From $\mathcal{F}$ we can then compute $\Im\{\mathcal{A}_0^* \mathcal{A}_1\}$
as\footnote{In the phenomenological scenario the Feynman rules for Majorana
neutrinos include spinors, charge conjugation and projection operators which we
assume to be included in $F_{\gtrless}$.}
\bigformula{
	\Im\{\mathcal{A}_0^* \mathcal{A}_1\} =  -
\Im\bigg\{\frac{i^{-1}\mathcal{F}}{g_1 g_2 g_3}\bigg\}\dend
}{contribution to absorptive part of amplitude}
where $g_1$, $g_2$ and $g_3$ stand for the generic couplings associated with the
three vertices
in the one-loop diagrams at Fig. \ref{fig:circling rules}.

\section{\label{PhysGhost}Physical and ghost fields}

In this section we briefly review the conventional calculation of the
\CP-violating parameter
in real-time thermal field theory. However, we use a notation
that is helpful to understand the ambiguities emerging there, and that can be
more easily
compared to the results from non-equilibrium field theory.

An obvious problem with the real-time formulation for the computation of
$n$-point functions   is that there are in general $2^n$ such functions which
differ in the types of the external vertices.  Historically the correct function
was considered to be the one with all external vertices of type-1 (physical). In
this case \eqn\eqnref{circling with fixed type of external vertices} leads to
the following formula for the imaginary part of a graph's contribution to the
amplitude:
\bigformula{
	\Im\left\{i^{-1}\mathcal{F}(1,\ldots ,1;z_j)\right\}
=\frac{1}{2}\sum_{\text{circling}\,(x_i),\, z_j}F_{\gtrless}(x_1 ,\ldots
,x_n;z_j)\kend
}{circling formula by Kobes Semenoff}
where the sum includes all possible circlings of the internal vertices $z_j$ but
only those circlings of external vertices $x_i$ which include both, circled and
uncircled vertices (indicated by the brackets around $x_i$).

The six diagrams contributing to the imaginary part of the three-point vertex
function according to \eqn\eqnref{circling formula by Kobes Semenoff} are shown
in \fig\figref{vertex corrections imaginary part by Kobes Semenoff}.
\begin{figure}[!ht]
	\centering
\includegraphics{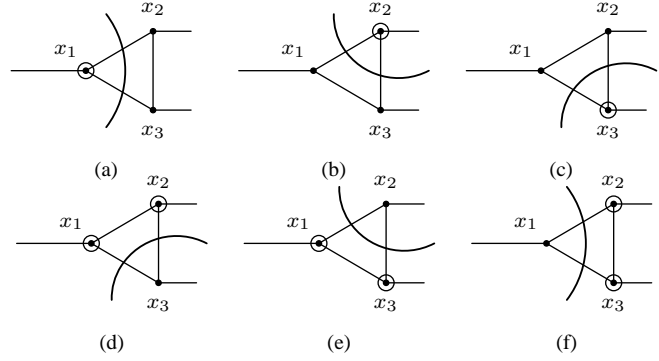}
		\caption{Circlings contributing to
$\Im\left\{i^{-1}\mathcal{F}(1,1,1)\right\}$ for the vertex loop. At one-loop
level the circlings can be interpreted as cuts, as indicated, by the lines
separating circled from uncircled regions \cite{Kobes1985np260p3}. The
contributions from diagrams involving cuts through the $x_2$-$x_3$ line are
suppressed relative to the others in the hierarchical limit.\label{fig:vertex
corrections imaginary part by Kobes Semenoff}}
\end{figure}
The circlings contributing to the self-energy part are shown in
\fig\figref{selfenergy loop corrections imaginary part by Kobes Semenoff}.
\begin{figure}[!h]
	\centering
\includegraphics{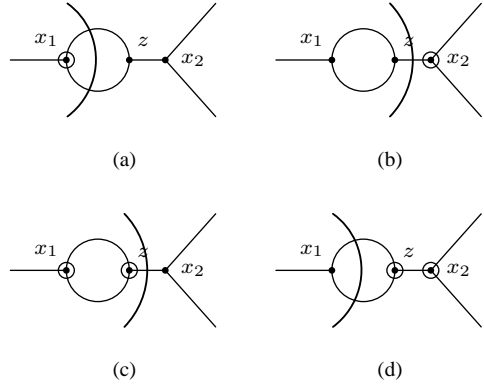}
		\caption{Circlings contributing to
$\Im\left\{i^{-1}\mathcal{F}(1,1;z)\right\}$ for the self-energy loop. The
graphs (b) and (c) vanish since $\psi_i$ and $\psi_j$ cannot be on-shell
simultaneously. Note that we consider only the diagrams with $\psi_i$ in the
external and $\psi_j$ in the internal line ($i\neq j$) because these are the
only ones which contribute to $\epsilon_i$.\label{fig:selfenergy loop
corrections imaginary part by Kobes Semenoff}}
\end{figure}

The contributions which correspond to cuts through the
$\psi_j$ line are suppressed in the hierarchical limit. If they are neglected
the application of this circling formula leads for the toy model to the result 
\bigformulalign{
	\cpviol{i}{V}{th}=-\frac1{8\pi}\frac{|g_j|^2}{\mpsii^2}\Im\left(\frac{g_i
g_j^*}{g_i^*
g_j}\right)\int\frac{\dif\Omega_l}{4\pi}\frac{1+\feq{\bbar}{E_1}+\feq{\bbar}{E_2
}+2\feq{\bbar}{E_1}\feq{\bbar}{E_2}}{
	\mpsij^2/\mpsii^2+\frac{1}{2}(1+\cos\theta_l)}+\ldots\kend
}{epsilon vertex thermal}
for the vertex contribution and 
\bigformulalign{
	\cpviol{i}{S}{th}=-\frac{\abs{g_j}^2}{16\pi}\Im\left(\frac{g_i g_j^*}{g_i^*
g_j}\right)\frac{1}{\mpsij^2
-\mpsii^2}\int\frac{\dif\Omega_l}{4\pi}\big\{{1+\feq{\bbar}{E_1}+\feq{\bbar}{E_2
}+2\feq{\bbar}{E_1}\feq{\bbar}{E_2}}\big\}
}{epsilon selfenergy loop thermal}
for the self-energy contribution. The distribution functions are to be evaluated
for the energies $E_{1}$ and $E_{2}$ given by\footnote{Note that if in \eqn\eqnref{epsilon 
selfenergy loop thermal} the term quadratic in the distribution functions was absent,
then by redefining the integration variable $\varphi_l$ we could write the energies 
$E_1$ and $E_2$ in the form 
$E_{{1,2}}={\frac12}\big[\e{\psi_1}{\momq}
+\momqabs(\sin\theta_l\cos\varphi_l\cos\delta' \mp
\cos\theta_l\sin\delta')\big]$, which was used in \cite{Hohenegger:2009a}.}
\begin{align}
	\label{IntermedEnergy}
	E_{{1,2}}={\frac12}\big[\e{\psi_1}{\momq}
\mp\momqabs(\sin\theta_l\cos\varphi_l\cos\delta' +
\cos\theta_l\sin\delta')\big]\kend
\end{align}
where $\theta_l$ and $\varphi_l$ are elements of the solid angle $\Omega_l$ and
the angle $\delta'$ is given in the limit of massless decay products by
$\sin\delta'=(\mompabs -\momkabs)/\momqabs$. The dots in \eqn\eqnref{epsilon
vertex thermal} represent further terms in $\feq{\psi_j}{}$ which are neglected.
Equivalently these results can be derived directly using only the $11$
components of the propagators, because the vertex and the self-energy loop do
not include internal vertices. Very similar results are known for the
phenomenological scenario which can be obtained using the propagators in
\eqn\eqnref{thermal Feynman rules fermions} for fermions for the Majorana
neutrinos and leptons in the loops and \eqn\eqnref{thermal Feynman rules
scalars} for the Higgs bosons \cite{Giudice:2004npb685,Covi:1998prd57}.
For the dependence on the distribution functions one obtains then the quadratic
form
\bigformula{
	1-\feq{\lepton}{E_1}+\feq{\higgs}{E_2}-2\feq{\lepton}{E_1}\feq{\higgs}{E_2}
\dend
}{thermal corrections to epsilon phenomenological scenario old}
The results obtained from non-equilibrium field theory in
\cite{Garny:2009rv,Garny:2009qn}
differ from \eqns\eqnref{epsilon vertex thermal} and \eqnref{epsilon selfenergy
loop thermal}. The
non-equilibrium results feature a different dependence on the distribution
functions,
\bigformula{
{1+\f{\bbar}{E_1}+\f{\bbar}{E_2}+2\f{\bbar}{E_1}\f{\bbar}{E_2} \quad \rightarrow
\quad  1+\f{\bbar}{E_1}+\f{\bbar}{E_2}}\dend
}{distribution dependence toy model}
Note that the top-down results are valid even if $\f{\bbar}{}$ is not an
equilibrium distribution ($\f{\bbar}{}\simeq \f{b}{}$ must hold, however) and
that the dependence is linear in the distribution function in contrast to
\eqns\eqnref{epsilon vertex thermal}, \eqnref{epsilon selfenergy loop thermal}.
The latter property  contradicts the result derived from thermal quantum field
theory .

In the phenomenological model, an analogous replacement
leads to a particularly important discrepancy. Indeed, \eqn\eqnref{thermal
corrections to epsilon phenomenological scenario old} would imply a cancellation
of the leading effects since
$\feq{\higgs}{p}-\feq{\lepton}{p}=2\feq{\higgs}{p}\feq{\lepton}{p}$.
The remaining effect is, in this case, entirely due to the fact that different
energies enter the distribution functions of leptons and Higgs particles in
\eqn\eqnref{thermal corrections to epsilon phenomenological scenario old}. Since
$E_1-E_2\sim \momqabs$, this  effect vanishes when the velocity of the Majorana 
neutrino in the medium rest-frame, $\momqabs/\e{N_1}{\momq}$,
becomes small.
Therefore it is important to check wether a replacement of the
form of \eqn\eqnref{distribution dependence toy model} does also occur in the
phenomenological scenario. This will be investigated in the next section.

\section{\label{Causal}Causal n-point functions}

We will now see how the finite temperature field theory approach can be
reconciled with the results derived from non-equilibrium quantum field theory.
In \cite{Kobes1990prd42,Kobes1991prd43,Eijck199plb278} it was shown that the
combination
\bigformulalign{
	\mathcal{F}_{R/A}^{(\alpha)}(x_1,\ldots ,x_n;z_j) = & \sum_{\text{circling}\,
x_i ,z_j}^{i\neq \alpha} F_{\gtrless}(x_1,\ldots x_\alpha ,\ldots ,x_n;z_j)\kend
}{retarded product}
referred to as the retarded (advanced) product, has the distinguishing property
that the time component $(x_\alpha)_0$ is singled out as being the largest
(smallest). This becomes clear when we consider the so-called largest (smallest)
time equation
\bigformula{
	F_{\gtrless} (x_1,\ldots ,x_\alpha ,\ldots ,x_n) + F_{\gtrless} (x_1,\ldots
,\underline{x}_\alpha ,\ldots ,x_n) =
0\,,\quad\text{if}\,\,(x_\alpha)_0\,\,\text{largest/smallest}\kend
}{largest smallest time equation}
which implies pairwise cancellation of the terms in \eqn\eqnref{retarded
product} if any external vertex $x_i$ with $i\neq\alpha$ has the largest
(smallest) time component. It has been realized that such causal products 
appear in Boltzmann equations in different cases, see for example \cite{Kobes1991prd43,
Weldon:1983}.
Furthermore, it has been shown that the causal products agree with the results
of the calculation in imaginary-time formalism analytically continued to real
energies, at least in a few examples including the self-energy loop and the
three-point vertex. 

The imaginary part of the causal sum was shown in \cite{Kobes1991prd43} to obey
\bigformulalign{
\Im\big\{i^{-1}\mathcal{F}_{R/A}^{(\alpha)}(x_1,\ldots ,x_\alpha ,\ldots ,
x_n;z_j)\big\} &=\nonumber\\
	\mp\frac{1}{2}\sum_{\text{circling}\, x_i}^{\text{not
all}}\sum_{\text{circling}\,z_j}^{}\Im\Big\{&i^{-1}F_> (x_1,\ldots
,\underline{x}_\alpha ,\ldots , x_n;z_j) - \nonumber\\
	- & i^{-1}F_< (x_1,\ldots ,\underline{x}_\alpha ,\ldots , x_n;z_j)\Big\}\kend
}{imaginary part of causal products}
where ``not all'' means that not all $x_i$ should be circled at the same time
and the imaginary part is taken of the causal product in momentum space. Here,
the vertex $x_\alpha$ with largest or smallest time is always circled. 

We can now compute the imaginary part of the advanced product
$\Im\big\{i^{-1}\mathcal{F}_{A}^{(1)}(x_1,x_2,x_3)\big\}$ for the three-point
vertex with smallest time component $(x_1)_0$ of the decaying particle. The
relevant circlings are shown in \fig\figref{vertex corrections imaginary part
causal products}. 
As before the contributions \fig\figref{vertex corrections imaginary part causal
products}(b) and (c) are suppressed due to the cut through the $\psi_j$
propagator line.
\begin{figure}[!ht]
	\centering
\includegraphics{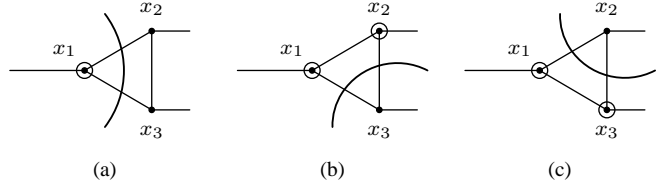}
		\caption{Circlings contributing to
$\Im\big\{i^{-1}\mathcal{F}_{A}^{(1)}(x_1, x_2 ,x_3 )\big\}$ for the vertex
loop. The advanced three-point function involves a difference of $F_>$ and $F_<$
contributions which differ by the replacement
$\Delta^{\pm}\leftrightarrow\Delta^{\mp}$ in the circling rules. Since the
finite temperature contributions (terms proportional to $f^{eq}$) to the latter
are the same, all contributions quadratic in the distribution functions cancel.
\label{fig:vertex corrections imaginary part causal products}}
\end{figure}

\newcommand{\projectorl}{P_L}
\newcommand{\projectorr}{P_R}
We compute the remaining contribution from \fig\figref{vertex corrections
imaginary part causal products}(a):
\bigformulalign{
	\Im\big\{i^{-1}\mathcal{F}_{A}^{(1)}(x_1, x_2
,x_3)\big\}=&\frac{1}{2}\Im\big\{i^{-1}F_{>}(\underline{x}_1,{x}_2,{x}_3)-i^{-1}
F_{<}(\underline{x}_1,{x}_2,{x}_3)\big\}\nonumber\\
	= & \frac{1}{2} \Im\Big\{i^{-1} \int \frac{\dif^4 l}{(2\pi)^4}
\Big[(+ig_1)(-ig_2)(-ig_3)D_{\beta}^- (q+l)D_{\psi_j} (q+l-k)D_{\alpha}^+ (l)
-\nonumber\\
 	& - (+ig_1)(-ig_2)(-ig_3)D_{\beta}^+ (q+l)D_{\psi_j} (q+l-k)D_{\alpha}^- (l)
\Big] S\Big\}\kend
}{causal circling formula contribution a 1}
where we take $F_{\gtrless}$ to include the spinors and charge conjugation
operators $C$ as well as projection operators $\projectorr,\,\projectorl$
associated with the vertices (for the Majorana neutrino interactions). This
leads to the trace part denoted by $S$ 
\cite{Buchmuller:1997yu,Giudice:2004npb685,Covi:1996wh}. In the massless lepton and Higgs
limit:
\bigformulalign{
	S=&\sum_{\text{spins}}\big[\bar{u}_{\lepton}(k)\projectorl
u_{N_i}(q)\big]^*\big[\bar{u}_{\lepton}(k)\projectorl (\gamma\cdot (q+l-k)
+M_j)\C^{-1} \projectorl (\gamma\cdot l) \projectorr\C
u_{N_i}(q)\big]\nonumber\\
	= &\tr\big[(\gamma\cdot q -M_i)(\gamma\cdot k)M_j (\gamma\cdot
l)\projectorr\big] \nonumber\\
	=&-2\,M_i M_j\, k\cdot l
}{trace part phenomenological}
and $S=1$ for the toy model.

It turns out that the pole of the $\psi_j$ propagator does not lie
in the loop integration region, so we can drop the $i\epsilon$ prescription. We
then get (the upper and lower signs correspond to the toy model and
phenomenological scenario respectively)

	\bigformulalign{
		\Im\big\{i^{-1}\mathcal{F}_{A}^{(1)}(x_1, x_2
,x_3)\big\}=-\frac{1}{2}\Im\int \frac{\dif^4 l}{(2\pi)^2} &  \delta\big((q+l)^2 
-\m{\beta}^2\big)\delta\big(l^2-\m{\alpha}^2\big)\frac{i}{(q+l-k)^2
-\mpsij^2}\times\nonumber\\
		\Big\{&\Big[\heaviside{-(q_0 +l_0)}\heaviside{l_0}-\heaviside{q_0
+l_0}\heaviside{-l_0}\Big]\nonumber\\
		\pm&\Big[\heaviside{-(q_0 +l_0)}-\heaviside{q_0
+l_0}\Big]\feq{\alpha}{l}+\nonumber\\
		+&\Big[\heaviside{l_0}-\heaviside{-l_0}\Big]\feq{\beta}{q+l}\Big\}S\kend
	}{causal circling formula contribution a 4}

which becomes
\bigformulalign{
	\Im\big\{i^{-1}\mathcal{F}_{A}^{(1)}(x_1, x_2 ,x_3)\big\}=\nonumber\\
	=\frac{1}{2}\int \frac{\dif^4 l}{(2\pi)^2} \delta\big((q+l)^2  &
-\m{\beta}^2\big)\delta\big(l^2-\m{\alpha}^2\big)\frac{1}{(q+l-k)^2
-\mpsij^2}\Big\{1\pm\feq{\alpha}{l}+\feq{\beta}{q+l}\Big\}S\dend
}{causal circling formula contribution a 5}
Performing the integration over $\dif \abs{\bvec{l}}\dif l_0$, this indeed leads
to the result for the \CP-violating parameter obtained in the top-down approach
with correct dependence on the distribution functions \eqn\eqnref{distribution
dependence toy model}. For the phenomenological scenario we obtain in the limit
of massless lepton and Higgs:
\bigformulalign{
	\epsilon_i^{V,th} &=\frac{1}{16\pi}\sum_{j\neq i}\frac{{\Im}\left\{(h^\dagger
h)_{ij}^2\right\}}{(h^\dagger
h)_{ii}}\frac{M_j}{M_i}\int\frac{\dif\Omega_l}{4\pi}\frac{1-\cos\theta_l}{{M_j^2
}/{M_i^2}+\frac{1}{2}(1+\cos\theta_l)}\big\{1-\feq{\lepton}{E_1}+\feq{\higgs}{
E_2}\big\}\kend
}{CP-violating parameter thermal vertex}
where $E_{1,2}$ are given by \eqn\eqref{IntermedEnergy}. 
In the zero temperature limit this reduces to the well-known result
\bigformulalign{
	\epsilon_i^{V,vac}&=-\frac{1}{8\pi}\sum_{j\neq i}\frac{{\Im}\left\{(h^\dagger
h)_{ij}^2\right\}}{(h^\dagger h)_{ii}}f\left(\frac{M_j^2}{M_i^2}\right)\kend
}{}
with 
\begin{align}
f(x)&=\sqrt{x}\left[1- (1+x)\ln\left(\frac{1+x}{x}\right)\right]\dend
\end{align}

The same computation can be performed for the self-energy loop. The possible
circlings are shown in \fig\figref{selfenergy loop corrections imaginary part
causal products}:
\begin{figure}[!ht]
\centering
\includegraphics{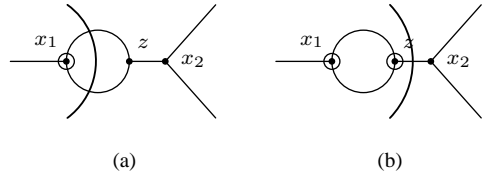}
	\caption{Circlings contributing to $\Im\big\{i^{-1}\mathcal{F}_{A}^{(1)}(x_1,
x_2 ; z)\big\}$ for the self-energy loop. Graph (b) vanishes since $\psi_i$ and
$\psi_j$ cannot be on-shell simultaneously.\label{fig:selfenergy loop
corrections imaginary part causal products}}
\end{figure}
\bigformulalign{
	\Im\big\{i^{-1}\mathcal{F}_{A}^{(1)}(x_1, x_2 ;
z)\big\}=&\frac{1}{2}\Im\big\{i^{-1} F_{>}(\underline{x}_1,{x}_2,{x}_3)-
i^{-1}F_{<}(\underline{x}_1,{x}_2,z)\big\}\nonumber\\
	=& -\frac{1}{2} \Im\Big\{ \int\frac{\dif^4 l}{(2\pi)^4}
\Big[(+ig_1)(-ig_z)(-ig_3)D_{\beta}^- (q+l)D_{\psi_j} (q)D_{\alpha}^+ (l)
-\nonumber\\
 	&\hspace{20mm} - (+ig_1)(-ig_z)(-ig_3)D_{\beta}^+ (q+l)D_{\psi_j}
(q)D_{\alpha}^- (l) \Big] S\Big\}\kend
}{causal circling formula contribution b 1}
where (the result for) $S$ coincides with \eqn\eqnref{trace part
phenomenological} in the phenomenological scenario, while $S=1/2!$ includes an
additional symmetrization factor in the toy model. This becomes
\bigformulalign{
	\Im\big\{i^{-1}\mathcal{F}_{A}^{(1)}(x_1, x_2 ; z)\big\}=\nonumber\\
	=\frac{1}{2}\int \frac{\dif^4 l}{(2\pi)^2} &  \delta\big((q+l)^2 
-\m{\beta}^2\big)\delta\big(l^2-\m{\alpha}^2\big)\frac{1}{q^2
-\mpsij^2}\Big\{1\pm\feq{\alpha}{l}+\feq{\beta}{q+l}\Big\}S\dend
}{causal circling formula contribution b 5}
This corresponds to the result for the self-energy contribution in the
hierarchical limit in the top-down approach \cite{Garny:2009qn} if the
equilibrium distribution functions are replaced with non-equilibrium ones. In
the zero temperature limit this leads to the correct vacuum result.
Thus, we have shown that (within the toy model) the \CP-violating parameter
$\epsilon^{th}$ obtained with help of thermal quantum field theory coincides
with the one obtained in the top-down approach (in the approximately symmetric
case) when one uses causal products instead of the conventional ones which
assume type-1 external vertices. Furthermore, by comparing with the top-down
result,
we find that the thermal field theory result can be generalized   to a
(symmetric) non-equilibrium configuration for the toy model  by the canonical
replacement of the equilibrium distribution functions with 
the non-equilibrium ones: $f^{eq}\rightarrow\f{}{}$.

\begin{figure}[!ht]
\centering
	\includegraphics[width=0.7\columnwidth]{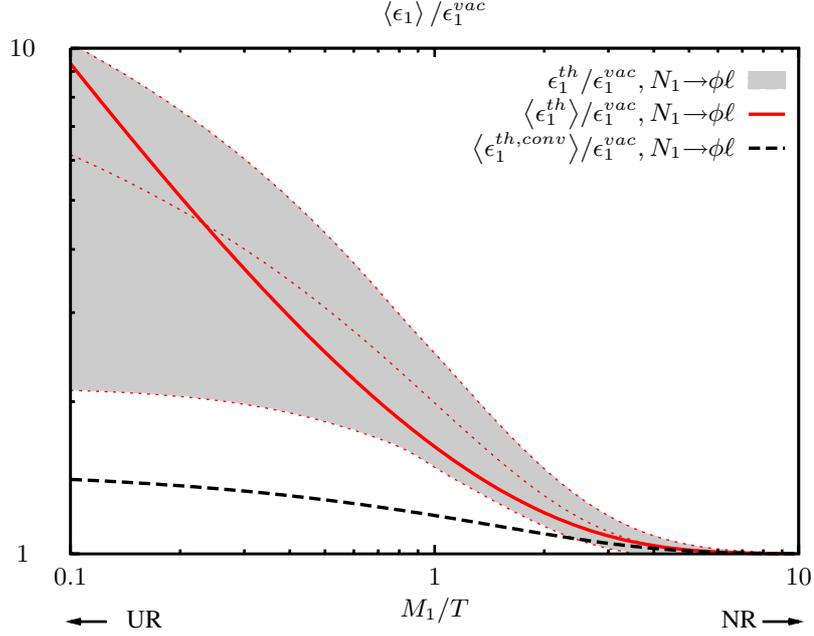}
	\caption{Temperature dependence of the \CP-violating parameter in the Majorana
neutrino decay relative to its vacuum value. Shown are the thermal average
$\langle\epsilon_1^{th}\rangle/\epsilon_1^{vac}$ (solid red line) and the values 
for various momentum modes $\epsilon_1^{th}/\epsilon_1^{vac}$ (dotted red lines) corresponding to
$\momqabs=T$, $-1\leq\sin(\delta')\leq+1$. For comparison we also show the 
conventional results $\langle\epsilon_1^{th,conv}\rangle/\epsilon_1^{vac}$
(dashed black line), where the leading effects cancel as described in the text.
Equilibrium distribution functions for bosons and fermions with negligible chemical potentials are
assumed. Note that the shown behavior can be modified if thermal masses are
included, since the decay $N_1 \rightarrow \lepton\higgs$ (and the conjugate
process) becomes kinematically forbidden if the thermal Higgs mass becomes too
large. At even higher temperature the process $\higgs\rightarrow N_1 \lepton$
becomes relevant instead \cite{Giudice:2004npb685}.\label{fig:epsilonMediumVsT}}
\end{figure}

\begin{figure}[!ht]
\centering
	\includegraphics[width=0.7\columnwidth]{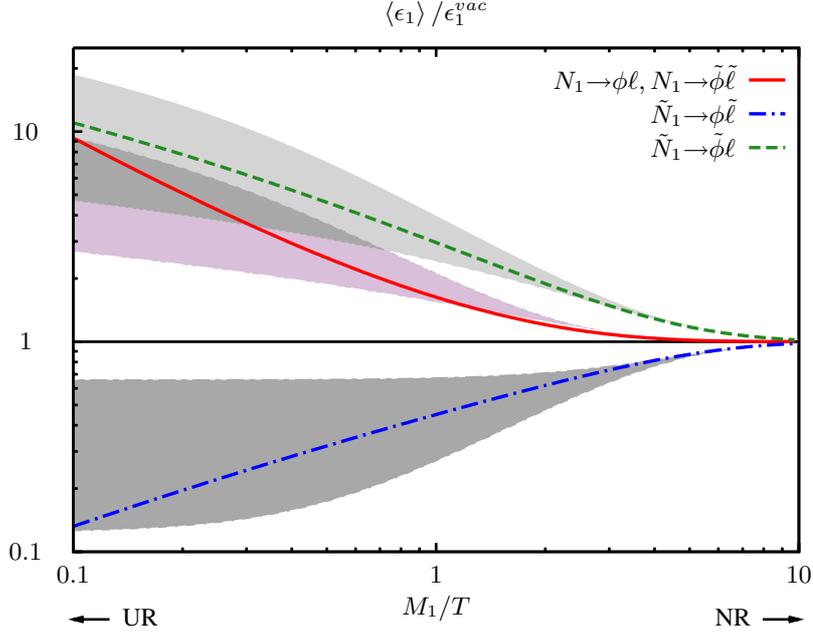}
	\caption{Medium correction to the \CP-violating parameters in the
        MSSM. The lines correspond to the thermal averages
        $\langle\epsilon_1^{th}\rangle/\epsilon_1^{vac}$, and the shaded
        regions illustrate the momentum dependence of
        $\epsilon_1^{th}/\epsilon_1^{vac}$ for $0.5\leq\momqabs/T\leq 4$
        and $\delta'=0$. Note also that in the weighted  sum of $\tilde N_1\rightarrow \phi {\tilde \ell}$
        and $\tilde N_1\rightarrow \tilde \phi {\ell}$ processes the 
        cancellation of the 
        medium contributions, observed in the earlier publications, does not occur
        anymore. 
       \label{fig:epsilonMediumVsT_MSSM}}
        
\end{figure}

For the phenomenological scenario we obtain in the limit of massless lepton and
Higgs for the self-energy contribution (including a factor of 2, because the two
components of the lepton doublet can propagate in the self-energy loop for a
given transition)
\bigformulalign{
	\epsilon_i^{S,th} &=-\frac{1}{8\pi}\sum_{j\neq i}\frac{{\Im}\left\{(h^\dagger
h)_{ij}^2\right\}}{(h^\dagger h)_{ii}}\frac{M_i M_j}{M_i^2
-M_j^2}\int\frac{\dif\Omega_l}{4\pi}(1-\cos\theta_l)\big\{1-\feq{\lepton}{E_1}
+\feq{\higgs}{E_2}\big\}\kend
}{CP-violating parameter thermal self-energy}
where $E_{1,2}$ are again given by \eqn\eqref{IntermedEnergy}.
In the zero temperature limit this reduces to the standard result
\bigformulalign{
	\epsilon_i^{S,vac}&=-\frac{1}{8\pi}\sum_{j\neq i}\frac{{\Im}\left\{(h^\dagger
h)_{ij}^2\right\}}{(h^\dagger h)_{ii}}\frac{M_i M_j}{M_i^2 - M_j^2}\dend
}{}
The complete \CP-violating parameter  is given by
\bigformulalign{
	\epsilon_i^{th}&=\epsilon_i^{V,th} +\epsilon_i^{S,th}\kend
}{CP-violating parameter thermal complete}
where the vertex and self-energy contributions are given by
\eqns\eqnref{CP-violating parameter thermal vertex} and \eqnref{CP-violating
parameter thermal self-energy} respectively. Therefore the overall dependence on
the distribution functions (vertex and self-energy contribution) is given by
 \bigformula{ 1-\feq{\lepton}{E_1}+\feq{\higgs}{E_2}. }{} 
In contrast to previous findings \eqn\eqnref{thermal corrections to epsilon phenomenological scenario old}, this  does \textit{not} vanish 
in the limit when the Majorana neutrino decays at rest assuming  
massless $\lepton$ and $\higgs$. Therefore, it is qualitatively different from the conventional result. 
The new expression can lead to a significant enhancement of the \CP-violating
parameter, see \fig\figref{epsilonMediumVsT}. 
Similar formulas can be derived
for processes such as $\phi\rightarrow N_1 \lepton$ in the standard model (which
can become relevant at higher temperatures) or for similar MSSM processes
involving sneutrinos and sleptons. The size of the 
medium corrections depends primarily on the statistics of the particles in 
the loop, see \fig\figref{epsilonMediumVsT_MSSM}.

\section{\label{BoltzmannEq}Boltzmann equations}

We can  assume in addition that the structure of the {\BES} for the
phenomenological scenario is analogous to the one given in
\cite{Garny:2009rv,Garny:2009qn} with appropriate quantum statistical factors
for bosons and fermions respectively and appropriate symmetrization factors.
This defines the full set of {\BES} including medium corrections to the
\CP-violating parameter for the phenomenological scenario as derived above.
With these modifications, the minimal network of quantum corrected Boltzmann
equations for thermal leptogenesis with hierarchical Majorana neutrino masses
$M_1 \ll M_2 , M_3$ takes the form (in homogeneous and isotropic
Friedman--Robertson--Walker space-time and not writing equations for the Higgs
fields $\higgs$, $\bar{\higgs}$ which are considered to be in thermal
equilibrium):
	\begin{subequations}
		\label{QuantumBoltzmannEquations}
		\begin{align}
			\label{boltzmann equation psi}
			\lorig
[\f{\Neutrino_1}{}](\momkabs)&=\carrayorigkb{\Neutrino_1\leftrightarrow
{\lepton}{\higgs}}{\momkabs}{\f{\Neutrino_1}{},\f{\lepton}{},\f{\higgs}{}}
			+\carrayorigkb{\Neutrino_1\leftrightarrow \bar{\lepton}\bar{\higgs}
}{\momkabs}{\f{\Neutrino_1}{},\f{\bar{\lepton}}{},\f{\bar{\higgs}}{}}\kend\\
			\label{boltzmann equation b}
			\lorig
[\f{\lepton}{}](\momkabs)&=\carrayorigkb{\lepton\higgs\leftrightarrow
\Neutrino_1}{\momkabs}{\f{\lepton}{},\f{\higgs}{},\f{\Neutrino_1}{}}\kend\\
			\label{boltzmann equation bbar}
			\lorig
[\f{\bar{\lepton}}{}](\momkabs)&=\carrayorigkb{\bar{\lepton}\bar{\higgs}
\leftrightarrow
\Neutrino_1}{\momkabs}{\f{\bar{\lepton}}{},\f{\bar{\higgs}}{},\f{\Neutrino_1}{}}
\kend
			\end{align}
	\end{subequations}
	where the Liouville operator is given by 
\bigformula{
	\lioulong{\f{a}{}}{x}{k} = k^0\left(\frac{\partial }{\partial
t}-\abs{\bvec{k}}\hubblerate\frac{\partial }{\partial
\abs{\bvec{k}}}\right)\f{a}{}({\abs{\bvec{k}}})\dend
}{covariant derivative in Friedman Robertson Walker spacetime}
If the generated asymmetry is small, as we assume here, then
$\f{\lepton}{}\approx \f{\bar{\lepton}}{}$ and $\f{\higgs}{}\approx
\f{\bar{\higgs}}{}$. In this case the CP-violating contributions to the
right-hand side of \eqn\eqref{boltzmann equation psi} cancel out  and we obtain
	\begin{align}
		\label{collision term psi-bb and psi-bbar bbar}
		\carrayorigkb{\Neutrino_1\leftrightarrow
\lepton\higgs}{\momkabs}{\f{\Neutrino_1}{},\f{{\lepton}}{},\f{{\higgs}}{}}+ &
\carrayorigkb{\Neutrino_1\leftrightarrow \bar{\lepton}\bar{\higgs}
}{\momkabs}{\f{\Neutrino_1}{},\f{\bar{\lepton}}{},\f{\bar{\higgs}}{}}
\nonumber\\
		\simeq{\frac{1}{2}}\int & \lorentzd{\lepton}{\momp} \lorentzd{\higgs}{\momq}
(2\pi)^4 \delta({\momk}-{\momp}-{\momq})\ampabs{N_1}{\lepton\higgs}^2
(\momp ,\momq)\nonumber\\ 
		\times \Big(& \big\{
\qstatferm{\f{\Neutrino_1}{{\momkabs}}}\f{\lepton}{{\mompabs}}\f{\higgs}{{
\momqabs}}-\f{\Neutrino_1}{{\momkabs}}\qstatferm{\f{\lepton}{{\mompabs}}}\qstat{
\f{\higgs}{{\momqabs}}}\big\}\nonumber\\	
+&\big\{\qstatferm{\f{\Neutrino_1}{{\momkabs}}}\f{\bar{\lepton}}{{\mompabs}}\f{
\bar{\higgs}}{{\momqabs}}
-\f{\Neutrino_1}{{\momkabs}}\qstatferm{\f{\bar{\lepton}}{{\mompabs}}}\qstat{\f{
\bar{\higgs}}{{\momqabs}}}\big\}\Big)\kend
	\end{align}
where, as usual, the tree-level amplitude for the Majorana neutrino decay is given by
$\ampabs{N_1}{\lepton\higgs}^2 (p,q)=\ampabs{N_1}{\bar\lepton\bar \higgs}^2
(p,q) = 2(h^\dagger h)_{11}\, \momp\cdot\momq$.
The collision terms for the (inverse) decay of the heavy particle into a
$\lepton\higgs$ or a $\bar{\lepton}\bar{\higgs}$ pair explicitly contain the
\CP-violating parameter $\epsilon_1$ given in \eqn\eqnref{CP-violating parameter
thermal complete} but with the equilibrium distributions replaced by
non-equilibrium ones $\feq{}{}\rightarrow\f{}{}$:
	\begin{subequations}
		\label{collision_term_bbbarbbarb-psi}
		\begin{align}
			\label{collision term bb-psi}
			\carrayorigkb{\lepton\higgs\leftrightarrow
\Neutrino_1}{\momkabs}{\f{\lepton}{},\f{\higgs}{},\f{\Neutrino_1}{}}=
&{\frac{1}{2}}\int  \lorentzd{\lepton}{\momp} \lorentzd{\Neutrino_1}{\momq}
(2\pi)^4 \delta({\momk}+{\momp}-{\momq}) \ampabs{N_1}{\lepton\higgs}^2
(\momk ,\momp)[1+\epsilon_1(\momqabs)]\nonumber\\
			\times &
\big\{\qstatferm{\f{\lepton}{{\momkabs}}}\qstat{\f{\higgs}{{\mompabs}}}\f{
\Neutrino_1}{{\momqabs}}
-\f{\lepton}{{\momkabs}}\f{\higgs}{{\mompabs}}\qstatferm{\f{\Neutrino_1}{{
\momqabs}}}\big\}\kend\\
			\label{collision term bbar bbar-psi}
			\carrayorigkb{\bar{\lepton}\bar{\higgs} \leftrightarrow
\Neutrino_1}{\momkabs}{\f{\bar{\lepton}}{},\f{\bar{\higgs}}{},\f{\Neutrino_1}{}}
=&{\frac{1}{2}}\int  \lorentzd{\bar{\lepton}}{\momp}
\lorentzd{\Neutrino_1}{\momq} (2\pi)^4
\delta({\momk}+{\momp}-{\momq})\ampabs{N_1}{\lepton\higgs}^2 (\momk
,\momp)[1-\epsilon_1(\momqabs)]\nonumber\\
			\times &
\big\{\qstatferm{\f{\bar{\lepton}}{{\momkabs}}}\qstat{\f{\bar{\higgs}}{{\mompabs
}}}\f{\Neutrino_1}{{\momqabs}}-\f{\bar{\lepton}}{{\momkabs}}\f{\bar{\higgs}}{{
\mompabs}}\qstatferm{\f{\Neutrino_1}{{\momqabs}}}\big\}\dend
		\end{align}
	\end{subequations}
Note that the network of Boltzmann equations \eqref{QuantumBoltzmannEquations}
should be understood in the generalized sense: the transition amplitudes differ
from the usual perturbative matrix elements and do not have their symmetry
properties as was noted in \cite{Garny:2009rv,Garny:2009qn}. The structure of
the collision terms \eqref{collision_term_bbbarbbarb-psi} differs from the
conventional one. In particular, we did not include the processes
$\lepton\higgs\leftrightarrow \bar{\lepton}\bar{\higgs}$ explicitly, because the
collision terms for the processes $\lepton\higgs\leftrightarrow\Neutrino_1$ and
$\bar{\lepton}\bar{\higgs}\leftrightarrow\Neutrino_1$ do not suffer from the 
generation of an asymmetry in equilibrium. To obtain a consistent set of
equations in the canonical bottom-up approach we would need to subtract the RIS
part of the $S$-matrix element for the processes $\lepton\higgs\leftrightarrow
\bar{\lepton}\bar{\higgs}$. Note, however, that it may be necessary to include
the collision terms for $\lepton\higgs\leftrightarrow \bar{\lepton}\bar{\higgs}$
(derived in the top-down approach) in quantitative studies, because these can
violate {\CP} in general. Further scattering processes with top-quarks and
gauge-bosons can also give relevant contributions.
We note here that this result should be treated with care, because additional
new effects could arise when the phenomenological scenario is investigated in
the top-down approach. In addition, the applicability of the quasi-particle
picture can not be tested in the framework of thermal field theory. In
particular the results presented above will only apply in the hierarchical case
\cite{Garny:2009qn}.
The analysis of the resonant case requires the use of the Kadanoff-Baym
formalism,
which allows us to take into account the in-medium spectral properties of the
mixing fields.

\section{\label{Conclusions}Conclusions}
Inspired by a discrepancy between conventional results for the thermal
corrections to the {\CP} asymmetries in thermal leptogenesis and recent new
results from non-equilibrium quantum field theory we have reconsidered the 
calculation of the \CP-violating parameters based on  thermal quantum field
theory. We find that, if causal products are used in the computation of the
$n$-point functions, the results of both approaches can be brought into
agreement in the framework of a toy model. We conclude that causal $n$-point
functions must be used in the derivation of the \CP-violating parameter in the
phenomenological scenario as well. This leads to new expressions for the thermal
corrections to the  vertex and self-energy  \CP-violating parameters.
In contrast to the conventional results the thermal corrections do 
\textit{not} vanish  in the limit when the Majorana neutrino decays at rest assuming  
massless decay products. Therefore, it is qualitatively different from the conventional result and might give significant contributions to the
generated baryon asymmetry. In the range from $0.1$ to $10$ of the dimensionless
inverse temperature, thermal effects can enhance the \CP-violating parameter by
up to an order of magnitude. 
The asymmetry can be computed using
the  minimal set of Boltzmann equations for leptogenesis in SM+$3\nu_R$
presented here,
which are analogous  to the equations which have been derived earlier in the
framework of the toy model. These take into account  decays and inverse decays
and include all quantum statistical factors in a way which guarantees that no
asymmetry is generated in equilibrium. They can be applied in the case of
non-degenerate Majorana neutrino masses. For a detailed phenomenological
analysis it will be necessary to take into account further thermal effects such
as thermal masses and resummed thermal propagators as well as additional
\CP-violating processes which exist in phenomenological scenarios.

\subsection*{Acknowledgements}
\noindent
This work was supported by the ``Sonderforschungsbereich'' TR27 and by the
``cluster of excellence Origin and Structure of the Universe''. We would like to
thank J-S. Gagnon for useful discussions related to thermal quantum field
theory.

\gdef\fmfoverset#1#2#3#4{%
			\parbox{#1}{
				{#3}
				\hspace{-#1}
				\vspace{-#2}
				{#4}
		}
}

\end{document}